\documentclass[aps,twocolumn,showpacs,english,pra,superscriptaddress]{revtex4-1}
\usepackage[latin1]{inputenc}
\usepackage{amsmath}
\usepackage{graphicx}
\usepackage[english]{babel}

\batchmode
\newcommand{\beq}{\begin{equation}}
\newcommand{\eeq}{\end{equation}}
\newcommand{\ket}[1]{| #1 \rangle}
\newcommand{\bra}[1]{\langle #1 |}

\begin{document}

\title{Optical drive of macroscopic mechanical motion by a single two-level system}

\author{A. Auff\`eves}
\email{alexia.auffeves@grenoble.cnrs.fr}
\affiliation{CNRS and Université Grenoble Alpes, Institut Néel, F-38042, GRenoble, France}

\author{M. Richard}
\affiliation{CNRS and Université Grenoble Alpes, Institut Néel, F-38042, GRenoble, France}

\pacs{85.85.+j,42.50.-p,03.65.Aa}




\begin{abstract}
A quantum emitter coupled to a nano-mechanical oscillator is a hybrid system where a macroscopic degree of freedom is coupled to a purely quantum system. Recent progress in nanotechnology has led to the realization of such devices by embedding single artificial atoms like quantum dots or superconducting qubits into vibrating wires or membranes, opening up new perspectives for quantum information technologies and for the exploration of the quantum-classical boundary. In this letter, we show that the quantum emitter can be turned into a strikingly efficient light-controlled source of mechanical power, by exploiting constructive interferences of classical phonon fields in the mechanical oscillator. We show that this mechanism can be used as a novel strategy to carry out low-background non-destructive single-shot measurement of an optically active quantum bit state.
\end{abstract}

\maketitle

\section*{Introduction}
Coupling electromagnetic and mechanical degrees of freedom is a new quest of quantum physics. First experiments were based on the interaction between the electromagnetic field stored in a cavity and a moving mirror, in the so-called linearized regime where the state of both resonators can be described classically \cite{Karrai,Arcizet2006,Gigan}. Recently, it has become possible to realize a similar setup, with a genuinely quantum system \cite{Treutlein} in place of the cavity, like a single spin in a diamond Nitrogen Vacancy (NV) center \cite{Arcizet}, a superconducting qubit \cite{Lahaye, Armour}, or a semiconductor quantum dot \cite{Wilson-Rae}. In the case of a optically active quantum system, the influence of the mechanical motion on the fluorescence properties has been recently evidenced \cite{Yeo}. In this letter, we investigate the opposite mechanism, and theoretically show that the quantum emitter submitted to modulated optical drive can efficiently pump phonons, thus behaving like an opto-mechanical transducer at the single quantum level. Two different excitation schemes are considered, either based on ultra-short pulses of light or on a continuous driving field resonant with the emitter's transition. In both cases, when the drive is modulated at the mechanical frequency, a coherent phonon field builds up by constructive interference. This scenario is intrinsically different from earlier proposals for mechanical cooling \cite{Rabl,Erik} or phonon lasing \cite{Kabuss}, where the control of the phonon population is achieved by side-band optical excitation, and not by intensity modulation. Finally, we suggest a method inspired from electron shelving technique used in ions traps \cite{Wineland,Steane}, to implement quantum non-demolition (QND) readout of the quantum system state. This proposal is applied to realistic systems like a quantum dot embedded in a nanobeam and superconducting qubits.

\begin{figure}[t]
\includegraphics[width=\columnwidth]{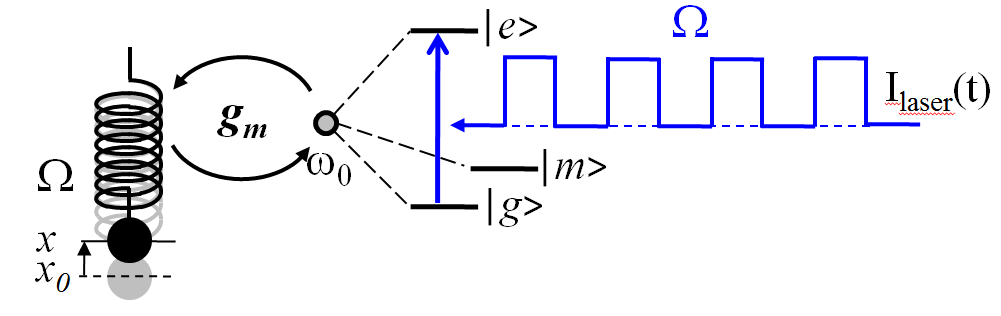}
\caption{Hybrid system under study: A quantum emitter (small gray circle in the center) with three possible states $\ket{g}$, $\ket{m}$, and $\ket{e}$ the excited state, is coupled to both the electromagnetic field and the motion of a macroscopic mechanical oscillator: the transition $\ket{g}$-$\ket{e}$ is directly coupled to light with a resonance frequency $\omega_0/2\pi$. It is also coupled with a strength $g_m$ to the position $x(t)-x_0$ of a macroscopic mechanical oscillator of rest position $x_0$, and of eigenfrequency $\Omega/2\pi \ll \omega/2\pi$. A laser beam, either consisting of ultrashort pulses, or continuous wave (CW), of intensity versus time $I_{laser}(t)$, is square-wave modulated at the frequency $\Omega/2\pi$ to modulate population in state $\ket{e}$. $\ket{m}$ is a dark state of the quantum emitter used for quantum non-demolition measurement.}
\label{fig1}
\end{figure}

\section{Optically-active quantum hybrid system}

The system under study is represented in Fig.\ref{fig1}. A quantum emitter interacts with a mechanical resonator while the population of the ground and excited states $\ket{e}$ and $\ket{g}$ are controlled by a modulated laser beam. The emitter-phonon coupling is modeled by the independent spin-boson Hamiltonian
\begin{equation}
H= \frac{\hbar\omega_0}{2}(\hat{\sigma}_z+1) + \frac{\hbar g_m}{2}(\hat{\sigma}_z+1)(b+b^\dagger) + \hbar\Omega b^\dagger b,
\label{hamiltonian}
\end{equation}
where $\omega_0/2\pi$ is the $\ket{g}$-$\ket{e}$ transition frequency, $\hat{\sigma}_z = \ket{e}\bra{e}-\ket{g}\bra{g}$ is the quantum emitter population, $b$ is the phonon annihilation operator, $\Omega/2\pi$ is the phonon frequency and $g_m$ is the coupling strength between the quantum emitter and the mechanical oscillator. Due to phonon thermalization and escape from the structure, the mechanical oscillator is subject to a damping rate $\Gamma$, while the quantum emitter exhibits a radiative relaxation rate $\gamma$. For what follows, we assume that the quantum emitter dynamics is much faster than that of the mechanical oscillator, i.e. $\omega_0,\gamma \gg \Omega$, such that it follows adiabatically the dynamics of the latter. In addition, since in general $\gamma \gg \Gamma$, the evolution of the system during a photon absorption/emission event can be considered Hamiltonian. Therefore, the Hamiltonian can be diagonalized in the subspaces $\ket{g}\bra{g}$ and $\ket{e}\bra{e}$ independently, yielding in the first case $H_g = \hbar \Omega b^\dagger b$ of eigenstates the usual Fock states $\ket{n}=b^{\dagger n}/\sqrt{n!} \ket{0}$, and in the second case, $H_e = \hbar (\omega_0 - g_m^2/\Omega) + \hbar \Omega B^\dagger B$, where $B=b+g_m/\Omega$. When the quantum emitter is excited, the new eigenstates are thus obtained by displacing the mechanical oscillator rest position by $d=-x_{zpf}g_m/\Omega$, where $x_{zpf}=\sqrt{\hbar/2m\Omega}$ is its zero point position fluctuation, and $m$ its effective mass \cite{Cleland}.

In order to provide a microscopic understanding of the system behavior under illumination, we first consider the effect of spontaneous emission of a photon by the quantum emitter on the mechanical state. The mechanical oscillator is initially described by a density matrix $\rho_0$. At $t=0$, an ultrashort pulse of light excites the quantum emitter into its state $\ket{e}$ (using e.g. a so-called "$\pi$"-pulse). Then for an average time $\tau=1/\gamma$ the quantum system remains in its excited state $\ket{e}$ and the evolution of the mechanical oscillator is governed by $H_e$. The system evolution operator from $t=0$ to $t=\tau$ thus reads $U_e(\tau) = e^{-iH_e\tau/\hbar}$. Once the photon is emitted, this evolution operator turns into $U_g(t) = e^{-iH_gt/\hbar}$ for $t>\tau$. The total evolution operator $U(t)=U_g(t-\tau)U_e(\tau)$ can be rewritten under the form $U(t)=U_g(t)D(-ig_m\tau)$ up to a phase factor, where $D(\alpha)$ is the displacement operator (by a complex number $\alpha$) in the complex plane. Thus, if $\rho_0=\ket{0}\bra{0}$ is the vacuum, any sequence of absorption/emission event generates a sum of tiny coherent phononic fields that interfere to build-up a coherent field $\ket{\beta(t)}$. If $\rho_0$ is a thermal state, the same sequence will simply displace it in the complex plane by $\beta(t)$. Therefore, since we are always dealing with one of these two initial states, the mechanical state will be well described at any time by the classical field $\beta(t)$, on top of some noise term: shot noise in the first case, and thermal noise in the second case.

This effect on the mechanical oscillator can be also understood in terms of jumps of the oscillator rest position: at $t=0^+$, the rest position $x_0$ jumps to $x_0+d$ so that the mechanical oscillator finds itself away from its rest position, and thus starts to move. After a typical time $\tau \ll 2\pi/\Omega$, when photon emission takes place, the rest position jumps back to $x_0$. This is the very mechanism which results in the mechanical excitation $\delta\beta$. Note that spontaneous emission is a stochastic process, such that the time $\tau$ spent in the excited state is a random quantity of mean value $1/\gamma$. As we will show later on, it means that the phonon gain $\delta\beta$ per pulse is also a fluctuating quantity.

\begin{figure}[t]
\includegraphics[width=0.9\columnwidth]{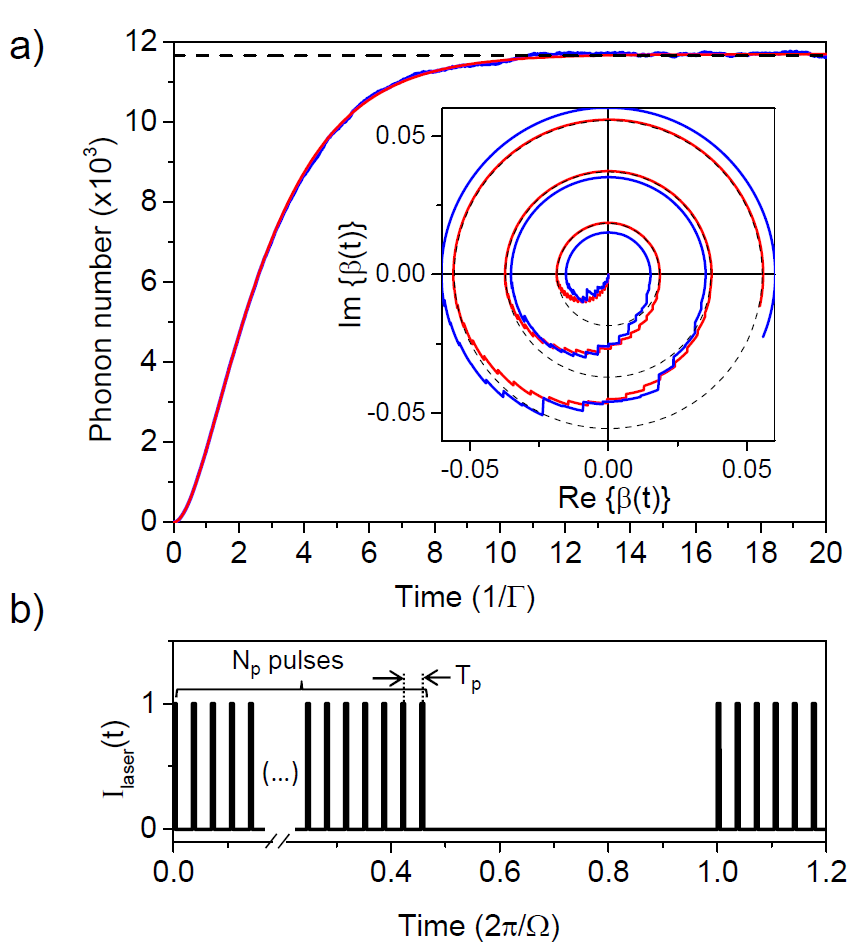}
\caption{(a) Phonon number $N$ versus time in units of $1/\Gamma$, with the initial condition $\beta(t=0)=0$. Solid red line, the excited state lifetime $\tau_{n,m}$ has been fixed to $1/\gamma$. Solid blue line, $\tau_{n,m}$ is a random quantity (cf. main text). The dashed line corresponds to the calculated $N_{max}$. Inset: $\beta(t)$ is plotted in the complex plane during the first three mechanical periods with $\tau$ fixed (red line) and random (blue line). The pumping session takes place in the lower quadrants, each pulse causing spike in $\beta$, resulting in a sawtooth pattern. The evolution is left free in the upper quadrants. To emphasize the pumping sessions, dashed circles of diameter $4n_pg_m/2\pi\gamma$ show free evolution of the field. (b) Time structure of the pulses trains used in this excitation scheme. Simulation parameters: $\Omega/2\pi=600$ kHz, $\Gamma=200$ Hz, $g_m/2\pi=450$ kHz, $1/\gamma=0.5$ ns. $T_p=41.7$ ns ($n_p=20$ pulses per mechanical period $2\pi/\Omega$).}
\label{fig2}
\end{figure}

\section{Modulated pulsed optical drive of the Two-level system}

$\delta\beta$ corresponds in general to a very small mechanical excitation. Thus, to increase it, we propose to use a train of optical pulses, of repetition rate $\gamma >f_p \gg \Omega/2\pi$ . For this strategy to be efficient, each microfield $\delta\beta$ generated by each pulse must interfere constructively with the phonon field already present in the mechanical oscillator. To do so, since $\delta \beta$ is a negative imaginary number, one must wait for the field $\beta$ already stored in the mechanical oscillator to be also negative imaginary. This condition is met every half-a-mechanical period when $\beta$ rotates within the lower quadrants of the complex plane. Therefore, to achieve constructive phononic interference, the pulse train must be square-wave modulated at the mechanical frequency $\Omega/2\pi$, with a $50\%$ duty-cycle (cf. Fig.\ref{fig2}.b).

There is a close analogy between this mechanical excitation strategy and the classical "forced oscillator regime" which is achieved, for instance, by hitting with a mechanical oscillator at a rate matching the mechanical eigenfrequency. However in our case, the hammer is microscopic, animated by light, and governed entirely by the law of quantum mechanics. This leads to a specific behavior of the coupled system as we point out in this work. In order to calculate the phonon field $\beta$ at all time under this modulated pulse trains excitation, we use the strategy described above for a single pulse, to derive recurrence relations (see supplementary materials for a detailed derivation):
\begin{align}
& \beta^m_{n+1} = (\beta^m_n-\frac{ig_m}{\Omega})e^{-i\alpha T_p}+\frac{ig_m}{\Omega}e^{-i\alpha(T_p-\tau_{n,m})} \label{pulsed1} \\
& \beta^{m+1}_{1} = \beta^m_{N_p}e^{-i\alpha\pi/\Omega}. \label{pulsed2}
\end{align}
Here $m$ refers to the $m^{\mathrm{th}}$ mechanical period, and $1\leq n \leq N_p$ refers to the $n^{\mathrm{th}}$ pulse among the $N_p=f_p/2\Omega$ pulses impinging on the quantum system during half a mechanical period. $T_p=1/f_p$, and $i\alpha=i\Omega+\Gamma/2$. Starting from an empty field, the phonon number $|\beta(t)|^2$ builds-up as shown in Fig.\ref{fig2}.a. After an elapsed time of a few $1/\Gamma$s, when the losses equal the pumping rate, the excited phonon number reaches its maximal value $N^p_{sat}=(2g_mf_p/\pi\Gamma\gamma)^2$. In the inset, the trajectory of the phonon field is shown in the complex plane. In the upper quadrants, the trajectory is circular as expected from free evolution with negligible losses. In the lower quadrants, the trajectory is spirale-like (cf. circular dashed lines for comparison) and features jumps which correspond to the constructive addition of microfields caused by the incoming pulses. For the red plots, a fixed value $\tau_{n,m}=\tau$ has been taken. For the blue plots the stochastic character of spontaneous emission has been taken into account by choosing $\tau_{n,m}$ at random in agreement with the probability rate $\rho(\tau_{n,m})=\gamma e^{-\gamma\tau_{n,m}}$. This stochasticity results in a fluctuating phonon gain per pulse which does not affect the overall dynamics as compared to the averaged value approach. This is expected since eq.(\ref{pulsed1}) shows that the gain per pulse $\delta\beta$ is directly proportional to $\tau$ as long as $\tau \ll 2\pi/\Omega$. Note that a finite quantum efficiency $\eta$ of the excitation can be taken into account in this model, due to e.g. excitation power too low to excite the quantum system with $100\%$ certainty for each pulse. It would just result in $N^p_{sat}$ decreased by a factor $\eta^2$. Note however that reaching $\eta\simeq 1$ poses no experimental difficulty if using for examples resonant $\pi$-pulses.

From a mechanical point of view, the quantum system serves as an opto-mechanical transducer which converts light into mechanical work. Due to its quantum nature, this conversion (input light power versus output mechanical power) is not linear and saturates to a finite mechanical power $P^p_{sat}=\hbar\Omega 4g_m^2/\pi^2\Gamma \times f_p^2/\gamma^2$. This power reaches a maximum value $P^p_{max}$ when the quantum system gets fully inverted by the pulse train, i.e. when $f_p=\gamma$:
\begin{equation}
P^p_{max}=\frac{\hbar\Omega 4g_m^2}{\pi^2\Gamma}.
\end{equation}
Note that increasing further $f_p$ has a detrimental effect, since when the system is still in state $\ket{e}$ it is either transparent to some incoming pulse, or in the case of $\pi$-pulses, it relaxes back into $\ket{g}$ before spontaneous emission occurs. These situations go beyond the limit of our model.

Interestingly, $P^p_{max}$ is quite large as compared to that produced by other proposed quantum size mechanical engines. Indeed, with realistic experimental parameters (cf. caption of Fig.\ref{fig2}) we find that $P^p_{max}=6.5\times 10^{-18} W$. This can be compared for example with a single trapped ion based heat engine undergoing an effective Otto cycle \cite{Abah}. There, a mechanical power of $10^{-20}$W is predicted.

\section{Modulated CW coherent optical drive of the Two-level system}
We now consider a second excitation scheme where the optical drive is provided by a continuous wave laser exactly resonant with the $\ket{g}-\ket{e}$ transition and modulated with a square-wave at the frequency of the phonon field $\Omega/2\pi$ with a $50\%$ duty cycle. For this excitation scheme, the interaction between the laser light and the quantum emitter needs to be included explicitly into the Hamiltonian (\ref{hamiltonian}) under the form $\hbar g(\hat{\sigma}_+e^{-i\omega_Lt}+\hat{\sigma}_-e^{i\omega_Lt})$, where $\omega_L/2\pi$ is the laser optical frequency, $\hat{\sigma}_{-(+)}$ is the lowering (raising) operator of the quantum emitter and $g/2\pi$ is the classical Rabi frequency. Like we have shown in the first part, the phonon field is well described by a classical field represented by a complex number $\beta(t)$. We thus adopt a semi-classical approach and derive generalized Bloch equations to describe the dynamics of the tripartite system:
\begin{equation}
\begin{array}{l}
\dot{s}_z = -2ig (s^* - s) - \gamma (s_z+1)\\
\dot{s}=-i\left[\delta + g_m(\beta+\beta^*)\right] s + \frac{ig}{2}s_z -\frac{\gamma}{2} s\\
\dot{\beta} = -i\Omega \beta -\frac{ig_m}{2}(s_z+1)-\frac{\Gamma}{2}\beta.
\end{array}
\end{equation}
We have defined $s_z = \langle \hat{\sigma}_z \rangle$, $s=\langle \hat{\sigma} \rangle$ and $\beta = \langle b \rangle$ the average population, dipole and phonon field amplitudes respectively. $\delta = \omega_0-\omega_L$ is the laser-emitter detuning when the oscillator is at its rest position. In the adiabatic regime considered here, the equations describing the evolution of the emitter and of the phonon field can be solved independently. At zero detuning, the emitter's population in the steady-state reads
\begin{equation}
s_z^\infty=-\left(1+\frac{g^2}{(\gamma/2)^2+[g_m(\beta_t+\beta^*_t)]^2}\right)^{-1},
\end{equation}
In general, the population of the excited state depends on the mechanically induced detuning $\delta'=g_m(\beta_t+\beta^*_t)$, that involves the phononic field phase and amplitude. However, in the saturated regime of the Bloch equations where $g \gg \gamma$, and in the limit $g \gg g_m|\beta_{max}|$, the excited state population is close to one half at any time and the equation of motion turns linear:
\begin{equation}
\dot{\beta} = -i\Omega \beta - i\frac{g_m}{2} - \frac{\Gamma}{2}\beta.
\label{CWdyn}
\end{equation}

\begin{figure}[t]
\includegraphics[width=0.9\columnwidth]{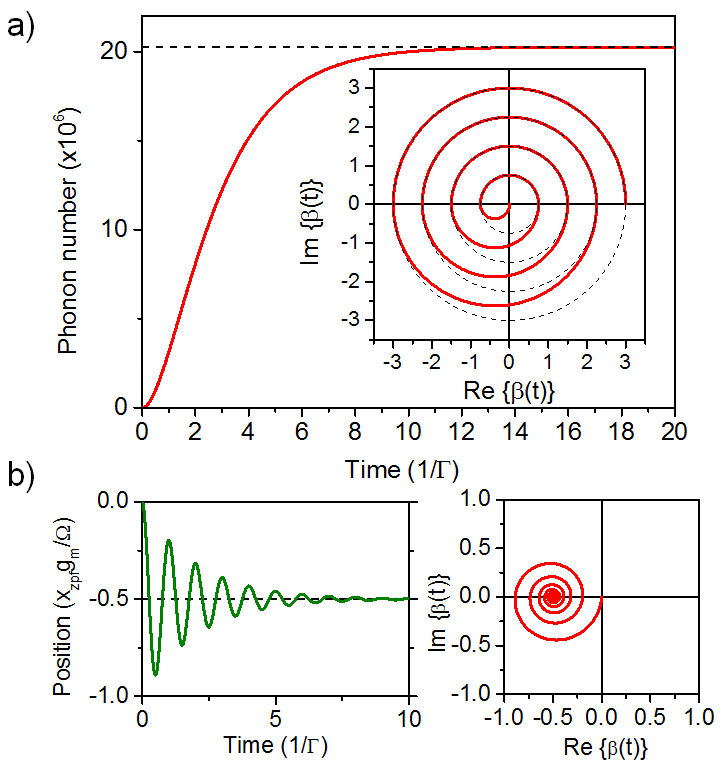}
\caption{(a) Phonon number $N$ versus time in units of $1/\Gamma$, starting from an empty phonon field at $t=0$. The dashed line corresponds to the calculated maximum phonon number $N^{cw}_{max}$. Inset: $\beta(t)$ is plotted in the complex plane during the first four periods with. The pumping session takes place in the lower quadrants. Free evolution of the field takes place in the upper quadrants. To emphasize the pumping sessions, dashed circles of diameters multiple of $\Delta\beta=g_m/\Omega$ show free evolution of the field. Simulation parameters: $\Omega/2\pi=600$ kHz, $\Gamma=200$ Hz, $g_m/2\pi=450$ kHz. (b) Left panel: Normalized position of the oscillator versus time under non-modulated CW resonant excitation (taking Q=2). The corresponding dynamics of $\beta$ is shown in the complex plane in the right panel.}
\label{fig3}
\end{figure}

The first term is the free evolution of the phonon field, the second one an effective phonon pumping term provided by the quantum emitter, and the last one describes the phonon damping. When the laser is on, the dynamics of the phonon field is governed by eq.(\ref{CWdyn}), while when the laser is off the phonon field evolves freely. From there, a new recurrent relation can be derived to express the phonon field $\beta_n$ after $n$ mechanical period (see supplementary information for details):
\begin{equation}
\left(\frac{g_m}{\alpha}+\beta_n e^{-1/4Q}\right)e^{-1/4Q}=\beta_{n+1},
\label{CWrec}
\end{equation}
where $Q=\Omega/2\pi\Gamma$. The build-up of the phonon field is shown in Fig.\ref{fig3} with the initial condition $\beta(t=0)=0$. The phonon number eventually saturates at $N^{cw}_{max}=\left|2Q g_m/\Omega\right|^2$, after a typical rise time (time to reach half-maximum phonon number) $T_r=2\ln(1-1/\sqrt{2})\Gamma^{-1} \simeq 2.46\Gamma^{-1}$. The mechanical power provided by the quantum emitter under this excitation scheme amounts to
\begin{equation}
P^{cw}=\frac{g_m^2}{\pi^2\Gamma}\hbar\Omega.
\end{equation}
Note that $P^{cw}$ does not depend on the laser excitation power. This is because the average excited state occupation does not depend on it either as long as it is large enough to meet the aforementioned conditions. When these conditions are not met, $P^{cw}$ will be lower either because large amplitude motion will detune the quantum emitter from the laser, or because too low laser intensity will prevent full Rabi oscillations to take place, which will result in average occupation of the excited state dropping below $0.5$. Note that $P^{cw}$ and $P^p_{max}$ only differ by a factor of $4$. This is expected since with the CW excitation scheme, the average excited state population, proportional to the phonon field at saturation, amounts to $0.5$ while in the pulsed scheme, it reaches $1$ for a repetition rate high enough. Therefore the latter strategy is more efficient, provided that one can tune the repetition rate of the laser at will.

Although both excitation strategies act on the system differently, they result in an identical behavior from the mechanical point of view at large timescale: owing to optical modulation, constructive phononic interferences result in the excitation of large amplitude mechanical oscillations. Note that shining unmodulated CW laser light on the quantum emitter does not result in phonons excitation. In fig.\ref{fig3}.b we have plotted the oscillator motion as a function of time in this condition. Right after switching on the laser, transient oscillations occur and the mechanical oscillator eventually stabilizes at a new position $x_0'$ shifted from its rest position $x_0$ by $x_{zpf}g_m/2\Omega$. This is a shift $2Q$ times smaller than the oscillation amplitude obtained in the modulated scheme. According to typical figures of realistic hybrid systems, this position shift is very small and requires at least the system to be in the so-called ultra-strong coupling regime \cite{Armour} to be measurable, i.e. $g_m/\Omega>1$.

\section{Application to quantum non-demolition measurements}
In spite of the quantum size of the opto-mechanical transducer, with our excitation schemes, the mechanical motion rise time $T_r$ can be fast and the amplitude impressively large, in particular as compared to the thermal phonon field. In the following we propose to exploit these properties to perform QND readout of a quantum state, using a technique inspired from electron shelving used in trapped ions experiments \cite{ions}. The two states of the system to be readout are $\ket{g}$ and $\ket{m}$ (see Fig.\ref{fig1}). Like previously, the quantum system is driven with a laser field resonant with the $\ket{g}-\ket{e}$ transition, modulated at the mechanical frequency $\Omega/2\pi$. If the quantum system is initially in the state $\ket{m}$, it is decoupled from the optical excitation and the phonon field remains in a thermal state of temperature $T$. If the quantum system is in $\ket{g}$, Rabi oscillations take place, pumping phonons in the mechanical mode. This large amplitude motion of the mechanical oscillator provides a good measurement of the quantum emitter state as soon as the pumped phonon number $N(t)$ exceeds the thermal one $N_{th}=(e^{\hbar\Omega/kT}-1)^{-1}$. Indeed, optical \cite{safavi} or electronic \cite{OConnell} measurement techniques have been demonstrated, with a measurement resolution of the mechanical oscillator position comparable with the zero point motion fluctuation.

In order to investigate the possibilities of this technique with state-of-the-art experimental parameters, we define the figure of merit $p=N(T_r)/N_{th}$, which compares the excited phonon population with the thermal one.
\begin{figure}[t]
\includegraphics[width=0.8\columnwidth]{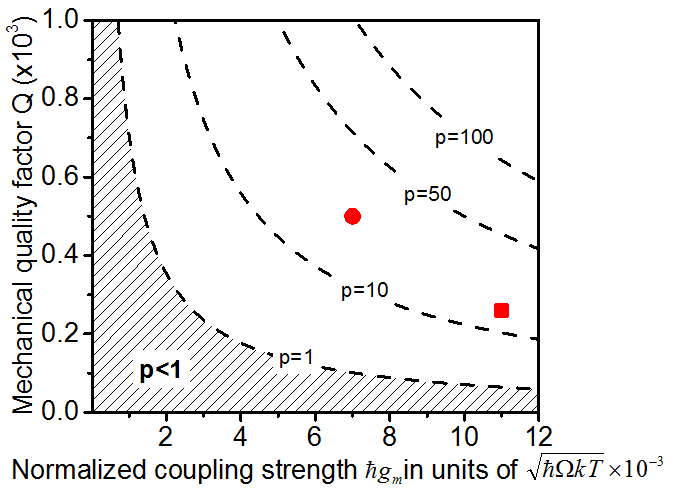}
\caption{Figure of merit $p$ (see main text) calculated at $t=T_r$ versus the coupling strength $\hbar g_m$ and the mechanical oscillator quality factor $Q$. The dashed lines are constant values of $p$. The symbols show the theoretical $p$ for two different realistic experimental configuration. The experimental parameters are $g_m/2\pi=62$MHz, $Q=260$, $T=25$mK, $\Omega/2\pi=6$GHz, for the superconducting circuit setup (square symbol). $g_m/2\pi=72$MHz, $Q=500$, $T=4.2$K, $\Omega/2\pi=1.2$GHz, for the semiconductor quantum dot embedded in a beam (circle symbol).}
\label{fig4}
\end{figure}
We calculate it for two different experimental realizations of light-coupled quantum hybrid systems. In the first one, a Josephson junction based qubit is coupled to a vibrating piezoelectric membrane \cite{OConnell}. Levels $\ket{g}$, $\ket{m}$ and $\ket{e}$ of our quantum emitter (cf. Fig.\ref{fig1}) can be identified with the first three lowest levels of the qubit arranged in a so-called $\lambda$-shape configuration \cite{Solano,Abdumalikov,Diniz}. As shown in Fig.\ref{fig4}, we find that $p=15$ is reached within the rise time $T_r=105$ns. Since the typical relaxation time of $\ket{g}$ and $\ket{m}$ are in the $50 \mu$s range \cite{Schuster}, measuring the mechanical oscillator motion versus time allows to perform projective readout of the qubit state in the single shot regime. An advantage brought up by this technique as compared to previously reported ones \cite{lupascu,vijay2,devoret} is that the probe and the signal are of different nature. Thus, the measurement can be carried out on a low background caused only by the thermal fluctuation of the mechanical motion.

The second system that we consider is a semiconductor quantum dot coupled to the motion of a nanobeam by the strain field \cite{Wilson-Rae,Yeo}. In this system, the three levels are arranged in a V-shape configuration, and the role of level $\ket{m}$ is played by the dark exciton state (total spin $J=2$) which lies slightly below the bright one. When the latter is occupied excitonic $\ket{g}$-$\ket{e}$ transition is quenched. Using the experimental parameters proposed in \cite{Wilson-Rae} we find a signal as large as $p=25$ within a rise time of $T_r=0.4 \mu$s. This is comparable with the dark exciton lifetime of typically $1\mu$s \cite{mcfarlane}, therefore quantum jumps between dark and bright exciton could be measured in real time with our technique. Note that the short lifetime of the bright exciton state (typically $1$ns) is not an issue here, since our measurement technique involves Rabi oscillations between $\ket{g}$ and $\ket{e}$ such that as long as $\ket{m}$ is not excited, $\ket{e}$ contains a fixed average population of one-half. Moreover recent developments have shown that the dark exciton state can be prepared optically \cite{poem}, and the bright exciton lifetime (typically $1$ns) can be increased to some significant extent using photoluminescence inhibition, as was demonstrated already \cite{ClaudonPRL} in a system which is directly usable as a quantum hybrid system \cite{Yeo}.

\section*{conclusion}
In this letter, we have shown that a single quantum emitter coupled to a macroscopic mechanical oscillator could be used as a very efficient opto-mechanical transducer. When its population is modulated at the mechanical frequency, large mechanical motion builds up by constructive phononic interference within the oscillator. This mechanism opens up interesting perspectives in terms of single shot readout of the quantum emitter internal degrees of freedom. This measurement regime has recently lead to ground-breaking demonstrations of quantum feedback and observation of quantum jumps with Rydberg atoms in microwave cavities \cite{Gleyzes2007,Sayrin} or Josephson qubits in circuit QED \cite{Vijay}. In our system, we show that single shot measurement of a QD state could be carried out with a high signal to noise ratio thanks to the different nature of the probe (optical) and the signal (mechanical). In addition, this mechanism could be exploited to manipulate the temperature of the mechanical mode: constructive interferences, as explained in this letter, allows large phonon field to build up. On the other hand, destructive interference could also be used to lower the mechanical resonance temperature. To do so, Brownian motion (characterized by random phase drift within the mode linewidth) needs to be measured in real time, and within a classical feedback loop, can be partially canceled by exciting a phonon field with the opposite phase. Note however, that such a method cannot be used to reach the phonon ground state as it is intrinsically classical, and that the final temperature would be limited by extrinsic parameters like the sensitivity of the Brownian motion measurement and the efficiency and bandwidth of the feedback loop. Finally, the results we have derived here have been obtained in the linearized limit of eq.(\ref{CWdyn}), i.e. $g$ larger than the oscillation amplitude of the quantum emitter energy. In the non-linear regime, this mechanism opens very interesting perspectives in the domain of quantum thermodynamics, where the laser plays the role of a heat bath for the quantum bit, while work is applied on it via mechanical motion \cite{Elouard}.

\begin{acknowledgments}
The authors thank P. Verlot, E. Dupont-Ferrier and O. Arcizet for fruitful exchanges. M.R. acknowledges the ERC starting grant "Handy-Q" n°258608, and A.A. acknowledges the support of the ANR JCJC "INCAL".
\end{acknowledgments}

\appendix
\section{Derivation of the recurrence relations in the situation of non-resonant $\Omega$-modulated pulsed excitation}

As explained in the main text, when a laser pulse hits the quantum system, the latter is assumed to be instantaneously excited from $\ket{g}$ to $\ket{e}$. This can be realized by setting the pulse resonant with an upper lying state efficiently coupled with $\ket{e}$, hence the term "non-resonant". Note that, even though coherent coupling between light and the quantum system is not required in this situation, the same result could be obtained by setting the pulse resonant with the $\ket{g}$-$\ket{e}$ transition and setting its "area" to $\pi$ (i.e. a pulse intensity that induces a $\pi$ rotation in the Bloch sphere from $\ket{g}$ to $\ket{e}$).

\begin{figure}[t]
\includegraphics[width=\columnwidth]{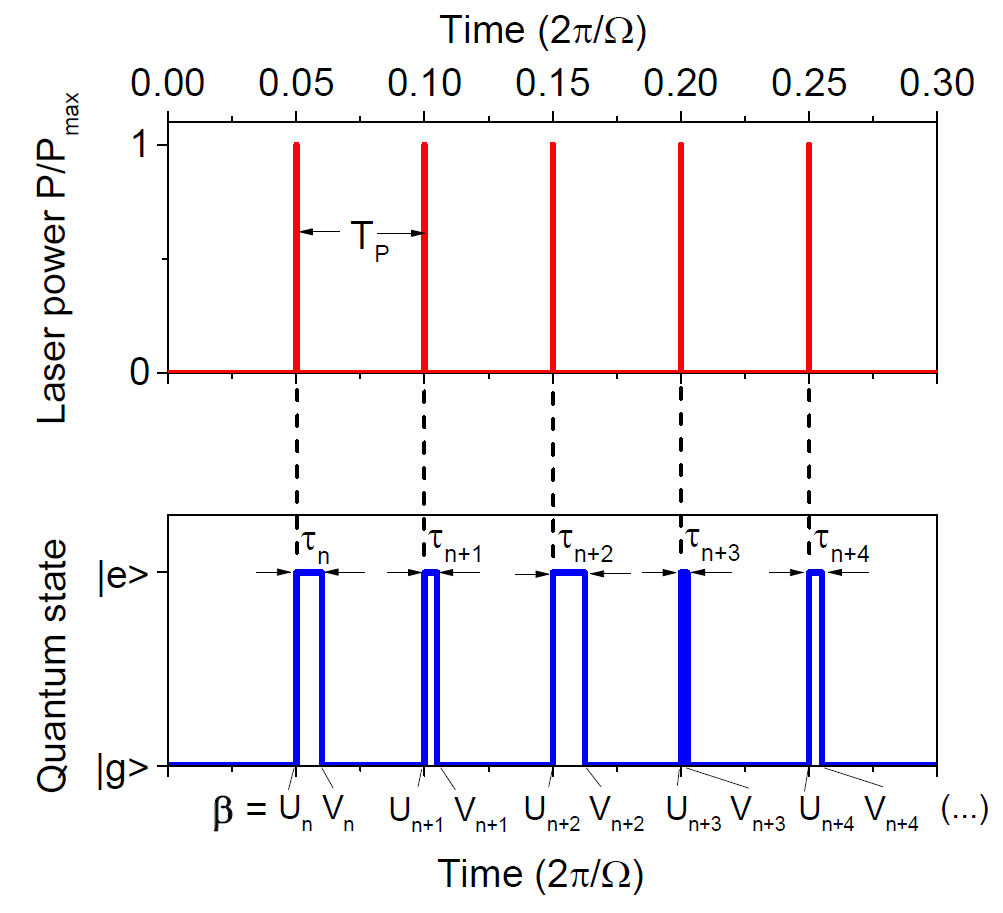}
\caption{Top plot: laser intensity versus time for pulsed excitation regime. The pulses are assumed delta-like and separated by a period $T_p$.
The state of the quantum system is shown in the plot below: when a pulse hits it, it is assumed to be instantaneously excited. Then it remains in its excited state $\ket{e}$ for an average time $\tau$ prior to radiative relaxation. the actual duration $\tau_n$ of the excited state fluctuates from pulse to pulse due to the stochasticity of spontaneous emission. According to the model described in the main text, the complex amplitude of the phonon field $\beta$ can be evaluated at specific times: $U_n$ is the field complex amplitude just before an incoming pulse, while $V_n$ is the field complex amplitude just after spontaneous emission has occurred.}
\label{Figs1}
\end{figure}

Once excited, this transition is subject to spontaneous emission and the system relaxes back into $\ket{g}$ after an average time $\tau$. This mechanism is illustrated in Fig.\ref{Figs1} where a few pulses of the pulse train are shown (upper plot), together with the dynamics of the quantum system state (lower plot). From the latter plot and the mechanism which couples the quantum system and the mechanical motion, as described in the main text, one can derive recurrent relations to calculate the mechanical complex amplitude $\beta$.

To do so, we exploit the fact that when the quantum system jumps into its excited state, $\beta$ is simply translated in the complex plane by a fixed real quantity $-g_m/\Omega$. Conversely, when it relaxes back to $\ket{g}$, $\beta$ is translated by an opposite quantity $g_m/\Omega$. Thus, the field $\beta=V_n$ is related to the field $U_n$ by performing first a translation in the complex plane, then a free evolution for a time $\tau_n$ and finally a translation back:
\begin{equation}
V_n=(U_n-g_m/\Omega)e^{-i\alpha\tau_n}+g_m/\Omega.
\end{equation}
Then $U_{n+1}$ can be related easily to $V_n$ (see Fig.\ref{Figs1} for its definition) as it corresponds to a free evolution of the field for a duration $T_p-\tau_n$:
\begin{equation}
U_{n+1}=V_ne^{-i\alpha(T_p-\tau_n)}.
\end{equation}
Combining these two equations results in eq.(2) of the main text. This recurrent relation allows one to describe the dynamics of $\beta$ for an uninterrupted sequence of pulses separated by $T_p$. However, as explained in the main text, in order to excite efficiently the mechanical motion, this pulse sequence must be modulated at the mechanical frequency $\Omega/2\pi$. To describe this pulse sequence, the index $n$ runs over a finite number $N_p$ of pulses: i.e. the number of pulses comprised within one-half of a mechanical period $2\pi/\Omega$ (See Fig.2.b of the main text). For the second half, due to the modulation, the laser is off and the field undergo a free evolution. Thus, at the very end of the mechanical period, the complex field amounts to
\begin{equation}
W=V_{N_p}e^{-i\alpha\pi/\Omega}.
\end{equation}
Finally, in order to have equations that express $\beta$ not only within a single mechanical period, the index $m$ needs to be introduced, which labels a given mechanical period. Doing so, we obtain eq.(3) of the main text.

\section{Derivation of the recurrence relations in the situation of resonant $\Omega$-modulated CW excitation}

In the main text, we show that the dynamics governing the phonon field with the quantum system under CW excitation is given by the differential equation (7). On the other hand, when the laser is off, the phonon field undergo free evolution :
\begin{equation}
\dot{\beta} = -i\Omega \beta - \frac{\Gamma}{2}\beta.
\label{free_ev}
\end{equation}
Eq.(7) has a general solution of the form $\beta_1=Ae^{-i\alpha t}-ig_m/2\alpha$, while eq.(\ref{free_ev}) has a solution of the form $\beta_2=Be^{-i\alpha t}$. Thus, in order to take into account the modulation, a strategy similar to the pulsed case is used: We start with an unknown field $\beta_0$ at time $t=0$. Let's assume that at $t=0^+$ the modulation phase is such that the laser is on, then $A_1=ig_m/2\alpha+\beta_0$. This solution is then propagated over the duration $T=\pi/\Omega$ of the "laser on" state (i.e. one half of the mechanical period). Thus at $t=T$, $\beta(T)=(ig_m/2\alpha+\beta_0)e^{-i\alpha T}-ig_m/2\alpha$. Then for $T<t\leq 2T$ the modulation switches off the laser for a duration $T$. Using eq.(\ref{free_ev}) and $\beta(T)=\beta(T^+)$ we derive $B_1=\beta(T)$ and $\beta(2T)=\beta(T)e^{-i\alpha T}$. by labeling with $n$ the $n^\mathrm{th}$ modulation period, this result reads like relation (8) of the main text.

\end{document}